# Technology-enhanced pre-instructional peer assessment: Exploring students' perceptions in a Statistical Methods course

Yosep Dwi Kristanto
Department of Mathematics Education, Universitas Sanata Dharma
Paingan, Maguwoharjo, Depok, Sleman, Yogyakarta 55282, Indonesia
E-mail: yosepdwikristanto@usd.ac.id

**Abstract**
There has been strong interest among higher education institution in implementing technology-enhanced peer assessment as a tool for enhancing students' learning. However, little is known on how to use the peer assessment system in pre-instructional activities. This study aims to explore how technology-enhanced peer assessment can be embedded into pre-instructional activities to enhance students' learning. Therefore, the present study was an explorative descriptive study that used the qualitative approach to attain the research aim. This study used a questionnaire, students' reflections, and interview in collecting student's perceptions toward the interventions. The results suggest that the technology-enhanced pre-instructional peer assessment helps students to prepare the new content acquisition and become a source of students' motivation in improving their learning performance for the following main body of the lesson. A set of practical suggestions is also proposed for designing and implementing technology-enhanced pre-instructional peer assessment.

**Keywords**: *peer assessment, pre-instructional activities, perceptions, Statistical Methods, higher education*

## Introduction

There has been strong interest among higher education institutions in implementing peer assessment as a tool for enhancing students' learning. Indeed, the growth of computer technology has a significant role in improving peer assessment applications in various educational settings (Yang & Tsai, 2010). It is also the case in mathematics learning. Mathematics education researchers have shown substantial evidence of technology-enhanced peer assessment's benefits on the students' learning (Chen & Tsai, 2009; Peter, 2012; Willey & Gardner, 2010). Specifically, Tanner and Jones (1994) posit that peer assessment helps the students to perform reflection through reviewing the works of others and recalling their own works.

Reflection process through which the students recall their existing mental context is fundamental components in learning (Lee & Hutchison, 1998; van Woerkom, 2010; Wain, 2017). Therefore, this process of reflection meets the purpose of pre-instructional activities. In the pre-instructional activities, it is expected that students can link their prior knowledge with the new content to be learned (Dick, Carey, & Carey, 2015). For this rationale, it is acceptable to stimulate reflection process by conducting peer assessment in pre-instructional activities. However, little has been shown in the literature that peer assessment is used in pre-instructional activities, though Scott (2017) has utilized the simulated peer assessment in improving numerical problem-solving skills as a prerequisite for learning Biology. The questions and the solutions used in Scott's study were not genuine students' works but were constructed by the researcher. Therefore, the present study tries to shed a light on how to embed technology-enhanced peer assessment into pre-instructional activities to enhance students' learning. This paper investigates students' perceptions in an attempt to portray students' learning.

Technology-Enhanced Peer Assessment

In understanding peer assessment, this study refers to the definition proposed by Topping (1998). He defined peer assessment as a process in which student measures the learning achievement of his/her peers. In the process, students have two different roles, namely assessors and assessees. As assessors, they evaluate and, in many cases, provide feedback to the works of their fellow students. In assessees role, they receive marking and feedback for their works and may act upon it.

Recent studies found that peer assessment has positive impacts on the students' learning. Several studies demonstrate that peer assessment can benefit the students in the assessment task, i.e. the quality of assessment they provided (Ashton & Davies, 2015; Gielen & De Wever, 2015; Jones & Alcock, 2014; Patchan, Schunn, & Clark, 2018). Furthermore, peer assessment also has effects on the students' acquisition of knowledge and skills in the core domain. In their study, Hwang, Hung, and Chen (2014) show that peer assessment effectively promotes the students' learning achievement and problem-solving skills. In particular, gaining learning achievement was also shown in Statistics class (Sun, Harris, Walther, & Baiocchi, 2015). One possible rationale of such benefits of peer assessment in the students' learning is the exposure to the works of their peers. When the students view their peers' works, they compare and contrast the works with their alternative solutions. This process of comparing and contrasting has the potential to facilitate students learning (Alfieri, Nokes-Malach, & Schunn, 2013; Reinholz, 2016).

Even though peer assessment has a number of advantages in facilitating learning, it also has several issues. The major concern in peer assessment is its validity as well as reliability (Cho, Schunn, & Wilson, 2006). Topping (1998) found disagreement on the degree of validity and reliability of peer assessment on his review, some studies report high validity and reliability (Haaga, 1993; Stefani, 1994; Strang, 2013), and the others report otherwise (Cheng & Warren, 1999; Mowl & Pain, 1995). However, the issues regarding validity and reliability can be reduced by providing the students with assessment rubrics (Hafner & Hafner, 2003; Jonsson & Svingby, 2007) since it makes expectations and criteria explicit.

Another issue regarding the peer assessment system is about administrative workload (Hanrahan & Isaacs, 2001). When implementing peer assessment in their class, instructors at least should manage the students' submission, assessment, and grading evaluation. Fortunately, these functions can be administered by using technology (Kwok & Ma, 1999). Technology can be used to record and assemble the results of scoring and commentary efficiently. In addition, technology also enables the teacher to provide immediate feedback based on the automated score calculation.

In the spirit of making the most of peer assessment's benefits and addressing its problems, peer feedback can be employed to accompany the peer assessment process. In peer feedback, the students discuss each other regarding performance and standards (Liu & Carless, 2006). They comment or annotate the draft or final assignments of their peers to give advice for the improvement of the assignments. When feedback comes with grading, it can be used to explain and justify the grade. It is also used to pose thought-provoking questions. The presence of the thought-provoking questions can foster the assessees' reflection on their assignments.

Pre-instructional Activity: Theory and Practice

From the instructional design perspective, Gagné, Briggs, and Wager (1992) posit that an instruction should be designed systematically to affect the students' development. Thus, instructional activities should be designed to facilitate the students' learning. One major component of the activities is pre-instructional activities. The activities are done prior to beginning formal instruction and it is significantly important to motivate the students, inform them the learning objectives, and stimulate recall of prerequisite skills. This study will not theoretically discuss all of the pre-instructional activities in depth. Instead, it will briefly present the examples of pre-instructional activities that appear in literature.

Pre-instructional activities can be done in different strategies. It also applies to mathematics learning. Loch, Jordan, Lowe, and Mestel (2014), in the Calculus of Variations and Advanced Calculus class, use screencasts to facilitate students in revising the prerequisite knowledge regarding the calculus techniques. Further, some scholars (Jungić, Kaur, Mulholland, & Xin, 2015; Love, Hodge, Corritore, & Ernst, 2015) use peer instruction as a pre-instructional strategy. The lesson introduction also can be done by simply telling the students of the prerequisites or testing them on entry skills (Conner, 2015).

**Method**

This study was an explorative descriptive research employing a qualitative approach in exploring how technology-enhanced peer assessment can be embedded into pre-instructional activities to enhance students' learning. The following sections give details of the research's setting, data collection, and also data analysis.

Research Setting

The research was conducted at a private university in Yogyakarta, Indonesia to investigate students' perceptions of the peer assessment system in Statistical Methods class. The class was conducted in a multimedia laboratory in which students have a computer to assist them in learning statistics. The author was the instructor of the class. The class utilized Exelsa, Moodle-based learning management system developed by the university, for the course administration purpose. In Exelsa, the students can access learning materials, post to a forum, and discuss with their peers about a certain topic, submit their assignments, assess and give feedback to their peers' works. The class was conducted biweekly with 24 meetings of instruction, one meeting of the midterm exam, and one meeting of the final exam. Each meeting consisted of 100-minute learning activities.

In three out of twenty-four meetings, the class was begun with peer-assessment activity. Therefore, students must submit their assignments before the class started. The assignments used in peer assessment were on the topics of one-way and two-way ANOVA. The assignments were done individually and required Microsoft Excel and SPSS Statistics in processing and analyzing real data given in the problems. The more details of the assignments will be described in the Findings section.

The peer assessment system used in this study was a workshop module (Dooley, 2009) provided by the LMS. The peer assessment takes place during the pre-instructional activities. The peer assessment system has five phases, i.e. setup, submission, assessment, grading evaluation, and reflection phases. In the setup phase, the instructor should set the introduction, provide submission instructions, and create an assessment form. After all of the components are set up, the instructor can activate the submission phase. In this phase, students can submit and edit their assignment. Optionally, they also can give a note on their assignments. However, students can only submit and edit their assignments before the class started.

Right after the class started, the instructor activated the assessment phase. In this phase, each student was assigned randomly to review assignments by their two peers. Thus, each student has two assessors. In reviewing their peers' assignments, students used a rubric to obtain a more objective assessment. The grading strategy used in the peer assessment system is the number of errors through which students grade each criterion by answering yes or no questions and optionally provide comments on the criterion. After all of the assignments were reviewed, the instructor can switch on the grading evaluation phase in which submission and assessment grade of each student were calculated automatically. In the end, students can directly see their score and feedback provided by their peers and reflect on it. The last mentioned is a reflection phase. The peer assessment process can be seen in Figure 1.

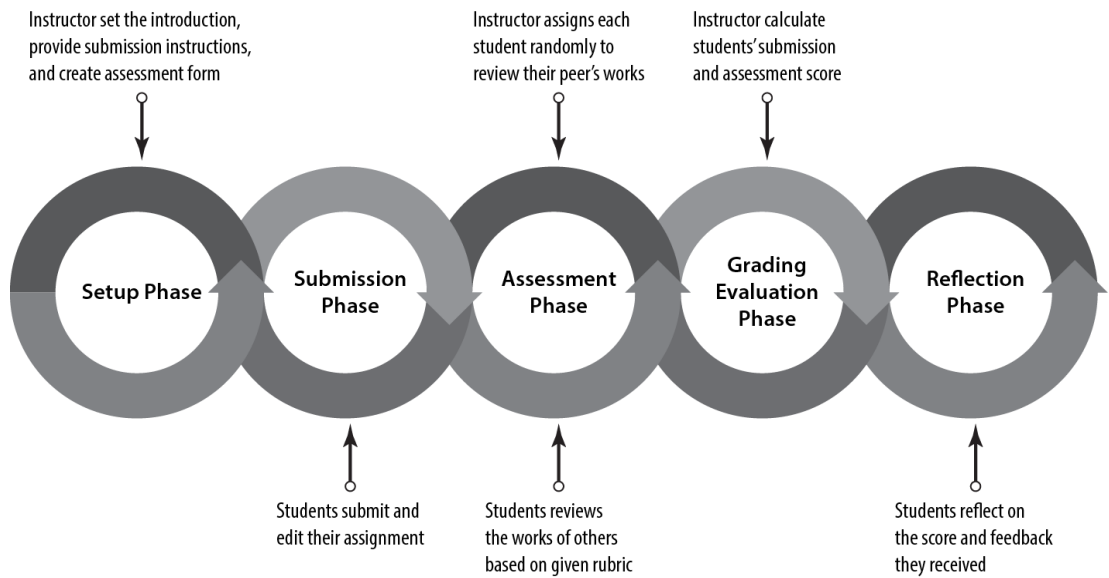

Figure 1. Peer assessment phases

Data Collection

The data collection process in this study was conducted between May and June 2018 and has been carried out in three phases. In the first phase, the researcher asked the students to write the reflection about their learning experiences in the course. The researcher prompted the students to use Gibbs reflective model (Gibbs, 1988). One learning experience that should be reflected by students was their experience in peer assessment activity. This phase of the data collection process was administered by the LMS.

In the second phase, a questionnaire adapted from Brindley and Scoffield (1998) was used to examine students' perceptions on the peer assessment. The questionnaire consists of three sections. The first section asked the students' personal data while the second section asked students' perceptions of peer assessment. The last section invited students to assess how useful the peer assessment process was. The second phase was done in the week right before the final exam and administered by Google form.

The third phase was conducted by interviewing three students on their general opinion about the learning process. The three students were purposively chosen to represent students' achievement. These students were interviewed simultaneously so they feel comfortable since the interviewer was their lecturer. The interview was recorded with the approval of these students to prevent data loss.

In addition, logs of three peer assessment activities in the LMS was also generated and downloaded. This logs file records the students' activities in the peer assessment system. Once downloaded, the logs data were then sorted in Microsoft Excel to know the duration of assessment task fulfillment done by each student. Moreover, the data also were used to find total time-frame of the assessment phase in each meeting.

Data Analysis

Data from the questionnaire and data logs were analyzed using descriptive statistics. Students' response from each item of the questionnaire was described as a proportion or mean value whereas data logs were described as a mean value for each meeting. Data from students' reflection and interview were examined and categorized by the researcher. The categories are derived from Wen and Tsai (2006, pp. 33–34) study, i.e. positive attitude, online attitude, understanding-and-action, and negative attitude. The data were labeled with the corresponding codes and analyzed via the Atlas.ti package program (for more information about conducting qualitative data analysis with Atlas.ti see, Friese, 2014).

Research Participants

In total, 34 students were enrolled in the author-taught course under study. Student gender demographics consisted of eight male and 26 female students. Most students were in their junior year with only five students from senior year. All of the students were prospective mathematics teachers.

**Findings and Discussion**

Findings

*Pre-instructional Activities Profiles*

The three meetings utilized technology-enhanced peer assessment in the pre-instructional activities. At the beginning of each session, the instructor informed the students about the learning objectives that should be achieved and linked the objectives with the previous assignments. The students were then asked to assess their peers' works through LMS. During the peer assessment process, the instructor moved about the classroom, observed students' progress on the assessment task, provided guidance if necessary, and answered questions if they arose. After the peer assessment process was complete, the instructor gave the students the opportunity to reflect on the score and feedback they received. The latter activity was the end of the pre-instructional activities.

The description of the assignments to be submitted before each meeting started is as follows. First meeting required students to submit an assignment on the topic of one-way ANOVA. The assignment asked students to investigate if there is a difference in the mean of football players' height in each position, i.e. forward, midfielder, defender, and goalkeeper. In the assignment, the instructor provided real data obtained from various sources. In this meeting, two students did not submit their assignment and there were also two students who submitted their assignment but did not attend the class.

In the second meeting, the students should have submitted a one-way ANOVA problem from the accompanying textbook (Bluman, 2012, p. 632). The problem asked them to determine the effective method in lowering blood pressure by examining the mean of individuals' blood pressure from three samples categorized by the methods they follow. The peer assessment process used in this meeting was slightly different from the previous meeting. In the assessment phase, the students had to assess an example submission provided by the instructor as an assessing practice before they assessed their peers' works. Three students did not submit their assignment in this meeting.

In the third meeting, students should have submitted their assignment for the 'Car Crash Test Measurements' problem from the accompanying textbook (Triola, 2012, p. 643). In this problem, the students were instructed to test for an interaction effect, an effect from car type and car size. One student did not submit their assignment in this meeting and there were also three students who submitted their assignment but did not attend the class.

The mean of assessment tasks carried out by all students in each meeting was calculated and reported in Table 1. On average, the period starting from the assessment phase begins until the assessment phase closes were 43.50 minutes. The table reveals that there has been a sharp decrease in the mean of first and second assessment tasks period carried out by the students in each meeting. In particular, the decreasing trend also applied in the second meeting when the students first reviewed an assessment example. In this meeting, the students reviewed example assessment in nearly a half of an hour (26.83 mins), the first peer's works in almost a quarter of an hour (12.69 mins), and the second peer's works in just over six minutes (6.72 mins).

Table 1. Mean of assessment phase time-frame in minutes

|  | Example assessment | Assessment 1 | Assessment 2 | Total |
|---|---|---|---|---|
| Meeting 1 | – | 20.03 | 5.83 | 39.97 |
| Meeting 2 | 26.83 | 12.69 | 6.72 | 59.80 |
| Meeting 3 | – | 10.77 | 4.83 | 30.72 |
|  | M = 26.83 | M = 14.50 | M = 5.79 | M = 43.50 |

*Students' Perception*

To investigate the students' perceptions of peer assessment, this study employed both quantitative and qualitative data. The quantitative data were obtained from the questionnaire, while the qualitative data were obtained from the students' reflections, the questionnaire, and interview.

From the questionnaire result, it is reported that most of the students (86.21%) in this study had previous experience on peer assessment. It is also found that approximately three out of four students perceived the necessity of assessing their peers. Further, it is only 27.59% of the students who fully understood the expectation imposed on them when reviewing their peers' works, whereas the rest only have a moderate understanding. In other words, all students understood what others expect on them in assessment tasks.

Four items of the questionnaire were rating-scale questions and used to explore the students' perceived easiness, fairness, pressure, and benefit of peer assessment. A mean report of the students' responses to the items is shown in Table 2. The students gave a high rating on fairness and responsibility of their marking (M = 4.07) and benefits of peer assessment they receive (M = 4.21). With regard to the grading task, they tend to posit that they have difficulties in assessing their peers' works (M = 3.24). However, they were under moderate pressure when they are doing the assessment task (M = 3.03). The sources of the pressure are various, more than half comes from their role (62.07%), almost a third comes from their experiences (31.03%), and the rest comes from their peers (6.90%).

The students' written reflection and interview are used to examine the students' perceptions as well. The perceptions were grouped into four defined categories and presented in Table 3. The main theme of the students' statements was the helpfulness of peer assessment in enhancing their learning. Regarding this theme, students stated that peer assessment helps them to enable reflective process, viz., reflecting on their mistakes shown by peers as well as reflecting and reviewing their own works to be compared and contrasted to peers' works. Second, the students perceived the peer assessment process as a tool for knowledge building since they should review their knowledge when assessing others. They added that assessing their peers encouraged them to discuss to their friends if they are indecisive about their assessment. This discussion led them to construct new knowledge to provide marking and feedback on the assessment task. Third, the students thought that peer assessment process develops their evaluative judgment making skills regarding their own works or others when they provide feedback to peers. Finally, the process of reviewing peers' works gives critical understanding and develops higher-level learning skills, such as analyzing and evaluating. The quotations from five students that reflect the benefits of peer assessment with regard to its usefulness in enhancing their learning are given below:

> *In my opinion, the peer assessment is useful. (It is) because it encourages me to review my own works if there is a mismatch between my own works and peers. So, (I) learned twice at once regarding the works. ($S_6$)*

> *… because I don't know (it is right or wrong) … I ask for help to my friend and found that my insight was improved. ($S_{15}$)*

> *This (peer) assessment was good to provide feedbacks to peers' works as well as to be responsible with my marking. ($S_{31}$)*

> *(Peer assessment) help us to think critically in assessing friends' works. ($S_{12}$)*

> *… we also must evaluate the answer of our friends which indirectly makes us reviewing the topics so that we can know/analyze where the friends' mistakes are. ($S_{29}$)*

Table 2. Students' perceptions scale on peer assessment

| Question | Mean |
|---|---|
| How difficult was assessing your peers' work? | 3.24 |
| How fair and responsible were you in assessing your peers' work? | 4.07 |
| How much pressure did the experience put you under? | 3.03 |
| How beneficial was the peer-assessment to you? | 4.21 |

Table 3. Categories of students' perceptions

| Category | Code (frequency) |
|---|---|
| Positive attitude | Helping learning (42) |
|  | Providing Feedback (5) |
|  | Enabling interaction (3) |
|  | Sustainability (3) |
|  | Helping instructor (2) |
|  | Engaging (1) |
|  | Motivating (1) |
| Online attitude | Anonymity (1) |
|  | Efficiency (1) |
|  | Transparency (1) |
| Understanding-and-action | Grading strategy (8) |
|  | Action for improvement (7) |
|  | Assessment criteria (2) |
| Negative attitude | Credibility (15) |
|  | No feedback (6) |
|  | Underestimating self-ability (3) |

Assessment credibility is another major theme of students' perceptions on peer assessment. On one hand, the students agreed that peer assessment gives the instructor other perspectives to provide more accurate grading and timely feedback. On the other hand, the students also questioned their peers' ability in assessing their works. It is possible that their peer assessors made an inaccurate assessment if the assessors' own works were inaccurate since the assessors often referred to it when undertaking an assessment task. Underrating self-ability also becomes a source of credibility issues. When the students feel incompetence on the subject-specific tasks, they are afraid of not being able to provide appropriate judgments. Reliability is students' next concern on peer assessment. They found that their assessors give different grades on the same item. Hence, they questioned peers' understanding of rubric criteria given by the instructor. The following are the students' statements related to the credibility of peer assessment.

> *Peer assessment is very useful as if the instructor makes an error on assessment, it can be remedied by peers' grading. ($S_8$)*

> *… However, the peer assessment doesn't work optimally when the assessor lacks understanding on what being assessed. (Moreover) the accuracy of each student's assessment is different from one another. ($S_{34}$)*

> *… Maybe the assessors' opinions are different from each other, since there are two friends that get different scores although their answers are more or less the same. ($S_{19}$)*

The students thought that feedback is an important component in peer assessment. Corrective feedback provided by peers was helpful for the students to know the errors on their works whereas suggestive feedback useful to make improvements later on. The importance of feedback was also reflected in students' responses when they did not receive feedback. They believe that assessors' task was not only give marking but also provide constructive comments. Some of the students' comments regarding the importance of feedback are as follows.

> *The one who said 'no' also comment. It is a constructive thing for us (to know) our mistakes that (the location of) the mistakes are in here, in here, and in here … There is a friend (that not only) said 'correct' but also give a comment, (you) should write like this and like this. So, that's the positive. It's like constructing (the understanding of) us. ($S_{15}$)*

> *Sometimes there is a friend who said that our answer was not correct, but does not give a single comment. That's it. So, we do not know where it goes wrong. ($S_{24}$)*

Other peer assessment aspects did not escape the students' attention. With regard to the number of errors grading strategy, they perceived that it provided not many options in marking peers' works. Instead of answering yes or no in each criterion, they prefer to use scale-rating strategy. However, they thought that the peer assessment process can facilitate students' discussion as well as students-instructor interaction. Other benefits of technology-enhanced peer assessment were also unfolded. Students stated that such assessment model was transparent and efficient as well as engaging and motivating.

Discussion

The aim of this study was to explore how technology enhanced peer assessment can be embedded into pre-instructional activities to enhance the students' learning. This paper interprets the students' perceptions in an effort to investigate students' learning experiences. In general, the research results show that technology-enhanced peer assessment holds significant promise to be an effective pre-instructional strategy. The learning benefits provided by peer assessment meet the purpose of the pre-instructional strategy.

The findings of the present study show that the process of assessing and commenting on the works of others facilitate the students' learning. This finding is in line with the result of prior studies in peer assessment investigation (Hanrahan & Isaacs, 2001; Sun et al., 2015). One possible explanation of this finding can be derived from comparative thinking perspective (Alfieri et al., 2013; Silver, 2010). When the student reviews peers' works, they compare and contrast it with their own works. If they doubt their own works, they ask for help to others or the instructor. This process of comparing and contrasting helps them to rehearse their own understanding that is useful for preparing them to gain new knowledge related to it.

The findings also suggest that peer assessment stimulates reflective thinking that drives action for improvement. Similar to the results of other studies (Davies & Berrow, 1998; Liu, Lin, Chiu, & Yuan, 2001), the peer assessment process leads the students to think critically and reflect the quality of their own works compared to the others'. This evaluative process helps the students to devise a plan in improving their learning products later on. As a feedback receiver, the students also take advantages of the feedback to enhance their learning. In other words, peer assessment can become a source of students' motivation in improving their learning performance in the commencing main body of a lesson (Jenkins, 2005).

The study also shows the importance of feedback in students' learning. As a salient element of peer assessment, peer feedback facilitates students in taking an active role in their learning (Liu & Carless, 2006). When the students provide corrective feedback on the peers' works, they develop an objective attitude in conducting their assessment task (Nicol & Macfarlane-Dick, 2006). Through providing suggestive feedback, the students think critically on the drawbacks of their peers' works even when the works are correct (Chi, 1996). As a feedback receiver, the students use peers' comment to improve their works. Moreover, peer's comments are potential to spark cognitive conflict when the comments contradict the student's prior knowledge. From the socio-cognitive perspective, cognitive conflict is fundamental in facilitating students' learning when it is successfully resolved (Nastasi & Clements, 1992).

However, the results of this study also reveal the resistance of peer assessment. Many students in this study have negative attitudes toward the fairness of peer grading. The similar result also can be found in the literature (Cheng & Warren, 1999; Davies, 2000; Liu & Carless, 2006). The negative perceptions come from the students' skepticism about the expertise of their fellow students. Even when a rubric was provided, the students thought that some of their peers were not really fair in giving marking. Another issue arose from grading strategy used in the assessment task. The correct and not-correct dichotomy into which students should categorize their peers' work is considered to be inflexible (Sheatsley, 1983). The students want more flexible grading strategy in order to be more confident in assessing their peers.

Last but not least, the study has several limitations to be considered. The first limitation of the current study relates to its exploratory design in investigating the students' learning experience. Future studies with a larger sample and a longer period are needed to verify the evidence found in this study. Second, this study only focuses on implementing peer assessment. Comparative studies are needed to compare the effectiveness of peer assessment and other strategies, such as advance organizers and overviews, to be used in pre-instructional activities. Finally, design-based studies could contribute to future literature in giving peer assessment design that optimizes the learning transition from lesson introduction to the main body of the lesson.

**Conclusion and Suggestions**

The contribution of this study is to show the potential of technology-enhanced peer assessment to be used as pre-instructional activities. The results of the current study, in general, suggest that the technology-enhanced pre-instructional peer assessment helps the students to prepare the new content acquisition for the following lesson. It is also found that peer feedback has a significant role in the peer assessment process in facilitating students' learning.

Based on the findings in the present study, the author proposed a set of suggestions for designing and implementing technology-enhanced pre-instructional peer assessment. First, a training should be provided to students so that they can provide and manage feedback as well as take action upon it effectively. Second, discussions between students and the instructor about assessment criteria are needed in order to improve students' understanding about what to be assessed by their fellow students' works. If necessary, the instructor also can invite students to develop the assessment criteria. Third, the instructor should monitor students' attitude toward grading strategy. This monitoring process aims to know the suitability of the grading strategy to students, tasks, and learning context. Finally, the instructor should use the assignment features (e.g., its content and context) used in peer assessment as a link to the commencing main body of the lesson.

**Acknowledgment**

The researcher would like to thank the students who participated in this study and LPPM of Universitas Sanata Dharma that supported this study. In addition, the researcher expresses gratitude to Russasmita Sri Padmi who kindly agreed to edit this manuscript.

**References**

Alfieri, L., Nokes-Malach, T. J., & Schunn, C. D. (2013). Learning through case comparisons: A meta-analytic review. *Educational Psychologist*, *48*(2), 87–113. https://doi.org/10.1080/00461520.2013.775712

Ashton, S., & Davies, R. S. (2015). Using scaffolded rubrics to improve peer assessment in a MOOC writing course. *Distance Education*, *36*(3), 312–334. https://doi.org/10.1080/01587919.2015.1081733

Bluman, A. G. (2012). *Elementary statistics: A step by step approach* (8th ed.). New York, NY: McGraw-Hill.

Brindley, C., & Scoffield, S. (1998). Peer assessment in undergraduate programmes. *Teaching in Higher Education*, *3*(1), 79–90. https://doi.org/10.1080/1356215980030106

Chen, Y., & Tsai, C. (2009). An educational research course facilitated by online peer assessment. *Innovations in Education and Teaching International*, *46*(1), 105–117. https://doi.org/10.1080/14703290802646297

Cheng, W., & Warren, M. (1999). Peer and teacher assessment of the oral and written tasks of a group project. *Assessment & Evaluation in Higher Education*, *24*(3), 301–314. https://doi.org/10.1080/0260293990240304

Chi, M. T. H. (1996). Constructing self-explanations and scaffolded explanations in tutoring. *Applied Cognitive Psychology*, *10*(7), 33–49. https://doi.org/

10.1002/(SICI)1099-0720(199611)10:7<33::AID-ACP436>3.0.CO;2-E

Cho, K., Schunn, C. D., & Wilson, R. W. (2006). Validity and reliability of scaffolded peer assessment of writing from instructor and student perspectives. *Journal of Educational Psychology*, *98*(4), 891–901. https://doi.org/10.1037/0022-0663.98.4.891.supp

Conner, K. (2015). Investigating diagnostic preassessments. *Mathematics Teacher*, *108*(7), 536–542.

Davies, P. (2000). Computerized peer assessment. *Innovations in Education and Training International*, *37*(4), 346–355. https://doi.org/10.1080/135580000750052955

Davies, R., & Berrow, T. (1998). An evaluation of the use of computer supported peer review for developing higher-level skills. *Computers & Education*, *30*(1), 111–115.

Dick, W., Carey, L., & Carey, J. O. (2015). *Systematic design of instruction* (8th ed.). Boston, MA: Pearson.

Dooley, J. F. (2009). Peer assessments using the moodle workshop tool. In *Proceedings of the 14th Annual ACM SIGCSE Conference on Innovation and Technology in Computer Science Education* (Vol. 41, pp. 344–344). New York, NY: ACM. https://doi.org/10.1145/1562877.1562985

Friese, S. (2014). *Qualitative data analysis with ATLAS.ti* (2nd ed.). London: SAGE.

Gagné, R. M., Briggs, L. J., & Wager, W. W. (1992). *Principles of instructional design* (4th ed.). Fort Worth, TX: Harcourt Brace College.

Gibbs, G. (1988). *Learning by doing: A guide to teaching and learning methods*. London: FEU.

Gielen, M., & De Wever, B. (2015). Structuring the peer assessment process: A multilevel approach for the impact on product improvement and peer feedback quality. *Journal of Computer Assisted Learning*, *31*(5), 435–449. https://doi.org/10.1111/jcal.12096

Haaga, D. A. F. (1993). Peer review of term papers in graduate psychology courses. *Teaching of Psychology*, *20*(1), 28–32. https://doi.org/10.1207/s15328023top2001_5

Hafner, J., & Hafner, P. (2003). Quantitative analysis of the rubric as an assessment tool: An empirical study of student peer-group rating. *International Journal of Science Education*, *25*(12), 1509–1528. https://doi.org/10.1080/0950069022000038268

Hanrahan, S. J., & Isaacs, G. (2001). Assessing self- and peer-assessment: The students' views. *Higher Education Research & Development*, *20*(1), 53–70. https://doi.org/10.1080/07294360123776

Hwang, G.-J., Hung, C.--Ming, & Chen, N.-S. (2014). Improving learning achievements, motivations and problem-solving skills through a peer assessment-based game development approach. *Educational Technology Research and Development*, *62*(2), 129–145.

Jenkins, M. (2005). Unfulilled Promise: Formative assessment using computer-aided assessment. *Learning and Teaching in Higher Education*, (1), 67–80.

Jones, I., & Alcock, L. (2014). Peer assessment without assessment criteria. *Studies in Higher Education*, *39*(10), 1774–1787. https://doi.org/10.1080/03075079.2013.821974

Jonsson, A., & Svingby, G. (2007). The use of scoring rubrics: Reliability, validity and educational consequences. *Educational Research Review*, *2*(2), 130–144. https://doi.org/10.1016/J.EDUREV.2007.05.002

Jungić, V., Kaur, H., Mulholland, J., & Xin, C. (2015). On flipping the classroom in large first year calculus courses. *International Journal of Mathematical Education in Science and Technology*, *46*(4), 508–520. https://doi.org/10.1080/0020739X.2014.990529

Kwok, R. C. W., & Ma, J. (1999). Use of a group support system for collaborative assessment. *Computers & Education*, *32*(2), 109–125.

Lee, A. Y., & Hutchison, L. (1998). Improving learning from examples through reflection. *Journal of Experimental Psychology: Applied*, *4*(3), 187–210. https://doi.org/10.1037/1076-898X.4.3.187

Liu, E. Z.-F., Lin, S. S. J., Chiu, C.-H., & Yuan, S.-M. (2001). Web-based peer review: The learner as both adapter and reviewer. *IEEE Transactions on Education*, *44*(3), 246–251. https://doi.org/10.1109/13.940995

Liu, N.-F., & Carless, D. (2006). Peer feedback: The learning element of peer assessment. *Teaching in Higher Education*, *11*(3), 279–290. https://doi.org/10.1080/13562510600680582

Loch, B., Jordan, C. R., Lowe, T. W., & Mestel, B. D. (2014). Do screencasts help to revise prerequisite mathematics? An investigation of student performance and perception. *International Journal of Mathematical Education in Science and Technology*, *45*(2), 256–268. https://doi.org/10.1080/0020739X.2013.822581

Love, B., Hodge, A., Corritore, C., & Ernst, D. C. (2015). Inquiry-based learning and the flipped classroom model. *PRIMUS: Problems, Resources, and Issues in Mathematics Undergraduate Studies*, *25*(8), 745–762. https://doi.org/10.1080/10511970.2015.1046005

Mowl, G., & Pain, R. (1995). Using self and peer assessment to improve students' essay writing: A case study from geography. *Innovations in Education and Training International*, *32*(4), 324–335. https://doi.org/10.1080/1355800950320404

Nastasi, B. K., & Clements, D. H. (1992). Social-cognitive behaviors and higher-order thinking in educational computer environments. *Learning and Instruction*, *2*(3), 215–238. https://doi.org/10.1016/0959-4752(92)90010-J

Nicol, D. J., & Macfarlane-Dick, D. (2006). Formative assessment and self-regulated learning: A model and seven principles of good feedback practice. *Studies in Higher Education*, *31*(2), 199–218.

Patchan, M. M., Schunn, C. D., & Clark, R. J. (2018). Accountability in peer assessment: Examining the effects of reviewing grades on peer ratings and peer feedback. *Studies in Higher Education*, *43*(12), 2263–2278. https://doi.org/10.1080/03075079.2017.1320374

Peter, E. E. (2012). Critical thinking: Essence for teaching mathematics and mathematics problem solving skills. *African Journal of Mathematics and Computer Science Research*, *5*(3), 39–43. https://doi.org/10.5897/AJMCSR11.161

Reinholz, D. (2016). The assessment cycle: A model for learning through peer assessment. *Assessment & Evaluation in Higher Education*, *41*(2), 301–315. https://doi.org/10.1080/02602938.2015.1008982

Scott, F. J. (2017). A simulated peer-assessment approach to improve students' performance in numerical problem-solving questions in high school biology. *Journal of Biological Education*, *51*(2), 107–122. https://doi.org/10.1080/00219266.2016.1177571

Sheatsley, P. B. (1983). Questionnaire construction and item writing. In P. H. Rossi, J. D. Wright, & A. B. Anderson (Eds.), *Handbook of survey research* (pp. 195–230). New York, NY: Academic Press.

Silver, H. F. (2010). *Compare & contrast: Teaching comparative thinking to strengthen student learning (A strategic teacher PLC guide)*. Alexandria, VA: Association for Supervision & Curriculum Development.

Stefani, L. A. J. (1994). Peer, self and tutor assessment: Relative reliabilities. *Studies in Higher Education*, *19*(1), 69–75. https://doi.org/10.1080/03075079412331382153

Strang, K. D. (2013). Determining the consistency of student grading in a Hybrid Business course using a LMS and statistical software. *International Journal of Web-Based Learning and Teaching Technologies (IJWLTT)*, *8*(2), 58–76. https://doi.org/10.4018/jwltt.2013040103

Sun, D. L., Harris, N., Walther, G., & Baiocchi, M. (2015). Peer assessment enhances student learning: The results of a matched randomized crossover experiment in a college statistics class. *PLoS ONE*, *10*(12), e0143177. https://doi.org/10.1371/journal.pone.0143177

Tanner, H., & Jones, S. (1994). Using peer and self-assessment to develop modelling skills with students aged 11 to 16: A socio-constructive view. *Educational Studies in Mathematics*, *27*(4), 413–431. https://doi.org/10.1007/BF01273381

Topping, K. (1998). Peer assessment between students in colleges and universities. *Review of Educational Research*, *68*(3), 249–276. https://doi.org/10.3102/00346543068003249

Triola, M. F. (2012). *Elementary statistics technology update* (11th ed.). Boston, MA: Addison-Wisley.

van Woerkom, M. (2010). Critical reflection as a rationalistic ideal. *Adult Education Quarterly*, *60*(4), 339–356. https://doi.org/10.1177/0741713609358446

Wain, A. (2017). Learning through reflection. *British Journal of Midwifery*, *25*(10), 662–666. https://doi.org/10.12968/bjom.2017.25.10.662

Wen, M. L., & Tsai, C.-C. (2006). University students' perceptions of and attitudes toward (online) peer assessment. *Higher Education*, *51*(1), 27–44. https://doi.org/10.1007/s10734-004-6375-8

Willey, K., & Gardner, A. (2010). Investigating the capacity of self and peer assessment activities to engage students and promote learning. *European Journal of Engineering Education*, *35*(4), 429–443. https://doi.org/10.1080/03043797.2010.490577

Yang, Y.-F., & Tsai, C.-C. (2010). Conceptions of and approaches to learning through online peer assessment. *Learning and Instruction*, *20*(1), 72–83.